\documentclass[twocolumn,nopacs,floatfix,superscriptaddress]{revtex4}
\usepackage{amsmath, amssymb}
\usepackage[utf8]{inputenc}
\usepackage{hyperref}
 \usepackage{xcolor}
\usepackage{graphicx}  
\newcommand{\eq}{\begin{equation}}
\newcommand{\eqx}{\end{equation}}
\newcommand{\eqn}{\begin{eqnarray}}
\newcommand{\eqnx}{\end{eqnarray}}

\begin{document}

\title{EEG Lempel-Ziv Complexity vis-à-vis Autocorrelation Function}

\author{Pedro Montoya}   
\affiliation{Cognitive and Affective Neuroscience and Clinical Psychology, Research Institute of Health Sciences (IUNICS) and Balearic Islands Health Research Institute (IdISBa), University of the Balearic Islands (UIB), Palma 07122, Spain }

\author{Ignacio Cifre} 
\affiliation{Universitat Ramon Llull, School of Psychology, Education and Sport Sciences Blanquerna, Barcelona 08022, Spain.}

\author{José Luis Rosselló Sanz} 
\affiliation{Artificial Intelligence Research Institute of the Balearic Islands (IAIB)
University of the Balearic Islands,(UIB), Palma 07122, Spain }
\author{Dante R. Chialvo}
\affiliation{Cognitive and Affective Neuroscience and Clinical Psychology, Research Institute of Health Sciences (IUNICS) and Balearic Islands Health Research Institute (IdISBa), University of the Balearic Islands (UIB), Palma 07122, Spain }
\affiliation{Artificial Intelligence Research Institute of the Balearic Islands (IAIB)
University of the Balearic Islands,(UIB), Palma 07122, Spain }
\affiliation{Center for Complex Systems and Brain Sciences (CEMSC3), Instituto de Ciencias F\'isicas (ICIFI-CONICET), Universidad Nacional de Gral. San Mart\'in, San Mart\'in, Buenos Aires, Argentina}
\affiliation{Consejo Nacional de Investigaciones Cient\'{\i}fcas y Tecnol\'ogicas (CONICET), Buenos Aires, Argentina}
\affiliation{Department of Physics, Faculty of Science, Hong Kong Baptist University, Hong Kong SAR, China}

\date{\today}

\begin{abstract}
Measures of signal complexity have become increasingly popular in electroencephalography (EEG), where they are often interpreted as indicators of neural information processing, dynamical richness, or conscious state. In contrast, classical measures of temporal dependence, such as the autocorrelation function (ACF), remain more directly linked to known physiological mechanisms and are easier to interpret. The present work revisits, for the EEG practicioner, the relation between normalized Lempel-Ziv complexity (LZC) and temporal autocorrelation in both synthetic and physiological EEG signals. First, synthetic time series exhibiting a broad range of temporal structure and complexity levels were generated using stochastic processes with varying degrees of temporal dependence. LZC and autocorrelation measures were then computed and compared. Across simulated signals, a robust inverse relationship was observed between normalized LZC and the first autocorrelation coefficient, indicating that signals with stronger temporal correlations exhibit lower complexity estimates. This relationship also holds in resting-state EEG recordings from healthy adult participants across channels and subjects, as well as dynamically during eyes open/close maneuvers. Overall, these results show that the information captured by LZC reflects temporal redundancy already quantified by conventional autocorrelation measures. Because autocorrelation measures relate more directly to established neurophysiological mechanisms, including oscillatory persistence, network synchronization, and temporal integration, these results suggest that interpreting complexity-based EEG findings should explicitly account for their dependence on conventional temporal statistics, which may provide a more transparent framework for mechanistic inferences.

\end{abstract}

\maketitle

\section{Introduction}
Measures of signal complexity have attracted growing attention in electroencephalography (EEG) research over the last two decades \cite{review1,review2,review3}. Complexity metrics have been applied to a wide variety of topics, including studies of consciousness, anesthesia, sleep, aging, neurodegeneration, epilepsy, and cognitive processing.  Due to its direct relationship with the notion of signal compressibility, Lempel-Ziv complexity (LZC) \cite{lzw,lzw2,lzw3} has emerged as one of the most widely used measures. In general terms, LZC quantifies the rate at which new patterns are encountered in a symbolic sequence. Less complex signals containing extensive repetition yield low complexity values because they can be compressed efficiently, whereas signals containing fewer repeated patterns yield higher complexity values.

 Numerous studies have interpreted changes in Lempel-Ziv complexity as evidence of altered information processing, changes in neural complexity, shifts in the dynamical repertoire, or differences in conscious state \cite{casali,schartner1,schartner2,schartner3}. In particular, reductions in LZC have been reported during anesthesia and non-rapid eye movement sleep, whereas increased LZC has been associated with psychedelic states and higher levels of conscious experience \cite{pasco}. The issue discussed in this note is that, despite its wide use in the literature, the basis of EEG Lempel-Ziv complexity measures remains unclear.

Although these findings show LZC’s sensitivity to changes in brain state, the signal properties underlying these changes remain incompletely understood \cite{ruffini}. Previous methodological work has emphasized that computing complexity estimates depend strongly on signal temporal correlations, spectral content, sequence length, symbolization procedures, and signal-to-noise ratio, complicating physiological interpretation \cite{aboy}. Furthermore, several authors have cautioned against equating changes in algorithmic complexity with changes in neural information processing or conscious content, noting that complexity measures often aggregate multiple statistical properties into a single index whose mechanistic meaning is not straightforward \cite{rapp,mediano2021,mediano}. Consequently, a given change in LZC may arise from alterations in oscillatory amplitude, spectral composition, temporal predictability, synchronization, or combinations thereof. 
This raises an important question: to what extent does LZC provide information beyond what simpler, more interpretable measures of temporal dependence, such as the autocorrelation function, already capture? The present study addresses this question directly.

\begin{figure*}[t!]
\centering

\includegraphics[width=0.8\linewidth]{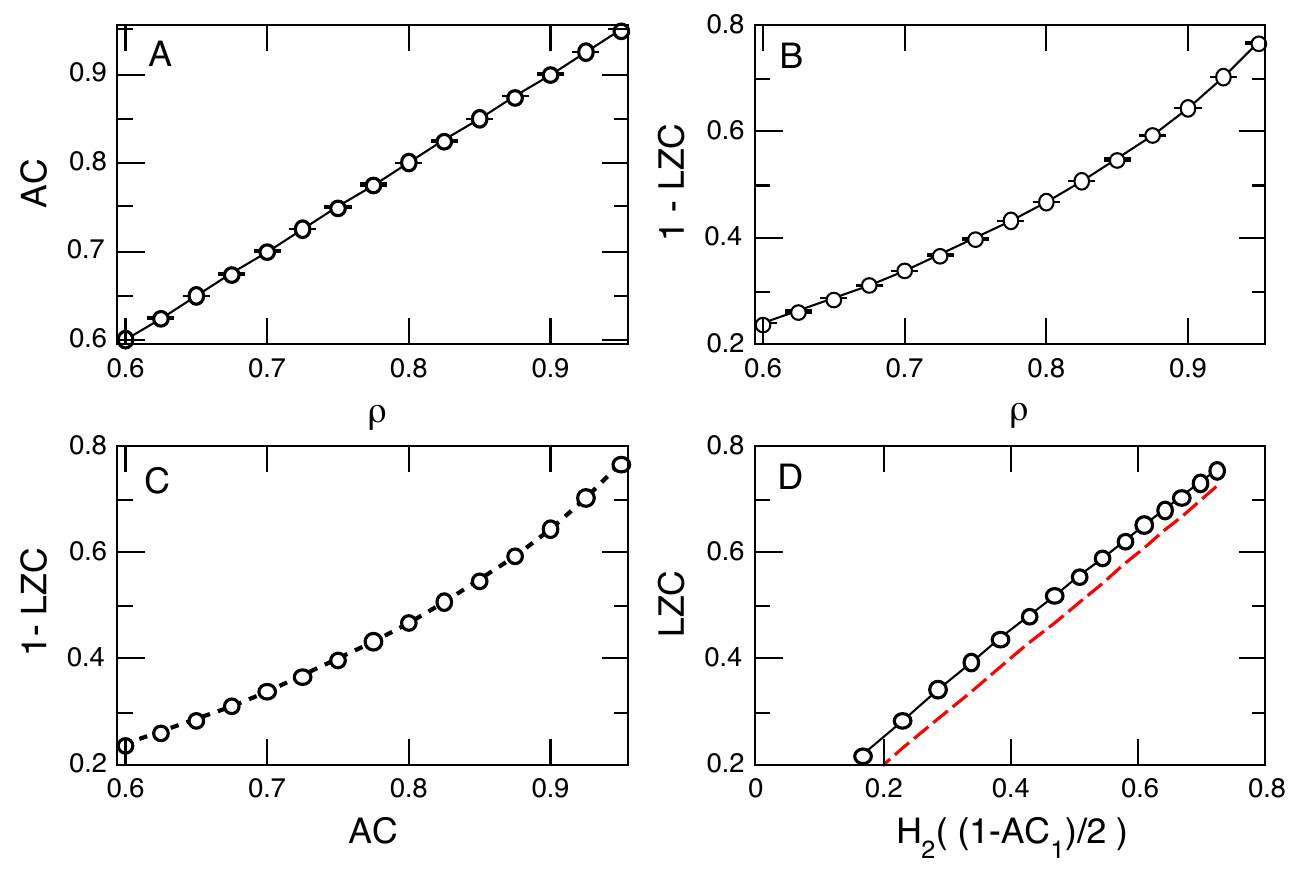}

\caption{Markov chain: Autocorrelation (AC) and Lempel-Ziv complexity (LZC) relation computed numerically. Panels A-C  show respectively, the numerically computed lag-1 autocorrelation (AC1),   Lempel-Ziv complexity (plotted as $1-\mathrm{LZC}$) across the range of the prescribed $\rho$ values, and the weakly non-linear association between AC and LZC. Panel D plots LZC against the binary entropy rate $H_2((1-AC_1)/2)$ obtained from the measured $AC_1$. The dashed red diagonal indicates the asymptotic reference $\mathrm{LZC}=h$. The systematic deviation of the finite-length LZC values from this diagonal is consistent with the known finite-size dependence of normalized Lempel-Ziv complexity, which converges very slowly to the entropy rate only in the infinite-length limit. As temporal dependence increases, the signal becomes more structured and predictable, as reflected in Lempel-Ziv complexity. The data points are means and SDs from $n=64$ realizations.}
\label{fig1}

\end{figure*}

In contrast with complexity metrics, autocorrelation analysis has a much longer tradition in signal processing and neuroscience. The autocorrelation (AC) function quantifies the similarity between a signal and time-shifted versions of itself. As such, it directly measures temporal dependence and average predictability. Time series with strong temporal persistence exhibit slowly decaying autocorrelation functions, whereas highly irregular or random signals display rapid loss of autocorrelation. In EEG research, autocorrelation has been associated with oscillatory persistence, synchronization processes, temporal integration, critical slowing, and the stability of large-scale network states.

At first glance, complexity and autocorrelation appear to represent fundamentally different concepts. Complexity is often viewed as an information-theoretic property, whereas autocorrelation is generally regarded as a statistical measure of temporal dependence. However, closer inspection suggests that the two quantities may be intimately related. The Lempel-Ziv algorithm estimates complexity by identifying repeated patterns within a sequence. Consequently, any process that increases temporal predictability should also increase pattern repetition and therefore decrease the estimated complexity. From this perspective, autocorrelation may constitute one of the principal determinants of LZC.

This note aims to illustrate and discuss this relationship systematically and is organized as follows: The next section defines the two metrics, LZC and AC.  In Section III, we use synthetic time series generated from stochastic processes with differing degrees of temporal dependence to investigate the association between these two metrics. In Section IV, we use physiological EEG recordings to test the relationship we uncovered and discuss its relevance to the brain criticality hypothesis \cite{chialvo2010}. Finally, we discuss the implications of these findings for interpreting EEG complexity measures and argue that autocorrelation often provides a more direct, physiologically interpretable framework for understanding the phenomena described by LZC.

\section{Metrics basic definitions}

For a discrete-time series $\{x_t\}_{t=1}^{N}$ (e.g., an EEG signal sampled at $N$ points), the autocorrelation function (ACF) measures the linear dependence of the signal with a delayed version of itself. The sample ACF at lag $k$ is defined as
\begin{equation}
    R_{xx}(k) = \frac{1}{N-k} \sum_{t=1}^{N-k} \left( x_t - \bar{x} \right) \left( x_{t+k} - \bar{x} \right) 
    \label{eq:acf}
\end{equation}
for $\quad k = 0, 1, \dots, K$;  where $\bar{x} = \frac{1}{N} \sum_{t=1}^{N} x_t$ is the sample mean and $K$ is the maximum lag considered (typically $K \ll N$). The normalized ACF is often reported as $\rho(k) = R_{xx}(k) / R_{xx}(0)$, so that $\rho(0)=1$. Equation~\eqref{eq:acf} quantifies how similar the time series is to a shifted copy of itself; for an EEG signal, a slowly decaying ACF indicates relatively strong short-term correlations, whereas a rapid decay suggests broadband or noisy activity.

To compute the Lempel-Ziv normalized complexity   of the same time series, one first converts the real-valued signal into a finite symbol sequence. A common approach is to binarize the series by comparing each sample to a threshold (e.g., the median or mean): $s_t = 1$ if $x_t \ge \bar{x}$, otherwise $s_t = 0$. This yields a binary string $S = s_1 s_2 \dots s_N$. The LZ complexity $c(S)$ is defined as the number of distinct substrings (or ``words”) encountered when parsing $S$ from left to right using the LZ76 algorithm \cite{lzw}: starting at the beginning, take the shortest prefix not seen before, then continue from the next symbol, and repeat until the end. Formally, if $S$ is parsed into $c$ blocks, then $c(S) = c$. For a binary string of length $N$, the normalized complexity is
\begin{equation}
    C_{\mathrm{LZ}} = \frac{c(S)}{c_{\max}(N)},
    \label{eq:lz}
\end{equation}
where the maximum complexity for a binary sequence of length $N$ is
\begin{equation}
    c_{\max}(N) = \frac{N}{\log_2 N}
    \label{eq:cmax}
\end{equation}
(up to a constant factor; often $c_{\max}(N) = N / \log_2 N$ is used as a practical upper bound). Thus $C_{\mathrm{LZ}} \in [0,1]$, with values near $1$ indicating high randomness (low compressibility) and values near $0$ indicating strong regularity (high compressibility). In addition,  $C_{\mathrm{LZ}}$  can be further normalized with the results from random shuffles of $S$.  For EEG analysis, $C_{\mathrm{LZ}}$ is typically computed on windowed segments (e.g., 1–10~s) to track changes in signal complexity over time.
 
{\bf \emph{Synthetic Time Series.}}
In order to simplify additional confounding factors, we generate binary sequences, obviating the process of binarization of the continuous time series. Thus, we consider two stochastic point processes, a Markov chain and a random Cantor set in 1D, and we also use a decimation process to progressively decorrelate a binary sequence.

{\bf \emph{Markov chain:}}
To demonstrate the relationship between complexity and temporal predictability under controlled conditions, we generated artificial time series spanning a broad continuum from highly unpredictable to highly persistent dynamics. First, we consider a discrete-time homogeneous Markov chain with two states, \(\{0, 1\}\), characterized by a symmetric transition probability matrix that depends on the autocorrelation parameter $\rho$.
The step-by-step transition dynamics between successive bits in the sequence are formally defined by the following conditional transition matrix equation:

\begin{align}
\mathbf{P} &= 
\begin{pmatrix} 
P(X_n = 0 \mid X_{n-1} = 0) & P(X_n = 1 \mid X_{n-1} = 0) \\ 
P(X_n = 0 \mid X_{n-1} = 1) & P(X_n = 1 \mid X_{n-1} = 1) 
\end{pmatrix} \notag \\[6pt]
&= 
\begin{pmatrix} 
\frac{1 + \rho}{2} & \frac{1 - \rho}{2} \\ 
\frac{1 - \rho}{2} & \frac{1 + \rho}{2} 
\end{pmatrix}
\label{Eq2}
\end{align}

To demonstrate the relationship between the two metrics, we generated synthetic signals with progressively increasing temporal dependence ($\rho = 0.6-0.95$). As seen in Figure \ref{fig1} (panel A), $AC1=\rho $ in the range explored, as expected analytically, since there is a perfect identity relationship between the parameter $\rho$ in Eq.\ref{Eq2} and the theoretical lag-1 autocorrelation coefficient AC1. Concurrently, Lempel-Ziv complexity varied systematically with increasing signal regularity (Figure \ref{fig1}, panel B), showing that the metric is highly sensitive to changes in the underlying temporal structure. Comparing AC1 with the complexity measure (plotted for visual convenience as 1 -LZ) shows that both indices follow a similar monotonic trend (Figure \ref{fig1}, panel C), supporting the notion that Lempel-Ziv-based measures capture information related to signal temporal organization.
 
For the symmetric binary Markov process considered here, the probability of changing state at each step is related to the theoretical lag-1 autocorrelation by
\begin{equation}
q=\frac{1-AC_1}{2}.
\end{equation}
When $AC_1$ is estimated from a finite realization, this relation holds approximately. The corresponding entropy rate is therefore
\begin{equation}
h(AC_1)=H_2\!\left(\frac{1-AC_1}{2}\right),
\end{equation}
where
\begin{equation}
H_2(q)=-q\log_2 q-(1-q)\log_2(1-q)
\end{equation}
is the binary entropy function. According to Ziv's theorem, for a stationary ergodic source, the normalized LZ76 complexity converges asymptotically to the entropy rate as the sequence length tends to infinity, although finite-length estimates may deviate from this limit \cite{estevezrams}. Therefore, a monotonic relationship between LZC and $H_2\!\left((1-AC_1)/2\right)$ is expected for sufficiently long sequences in this Markov model. This is illustrated in Fig.~\ref{fig1} (panel D), where the numerically estimated LZC exhibits a monotonic relationship with the entropy rate predicted from the measured $AC_1$, while remaining systematically above the asymptotic reference for the finite sequence lengths considered. Thus, in this simple Markov model, the observed relationship between temporal autocorrelation and Lempel--Ziv complexity is consistent with the analytical dependence of the entropy rate on $AC_1$, while the residual offset highlights the expected finite-size effects of the LZC estimator.

\begin{figure}[t!]
\centering
\includegraphics[width=	.95\linewidth]{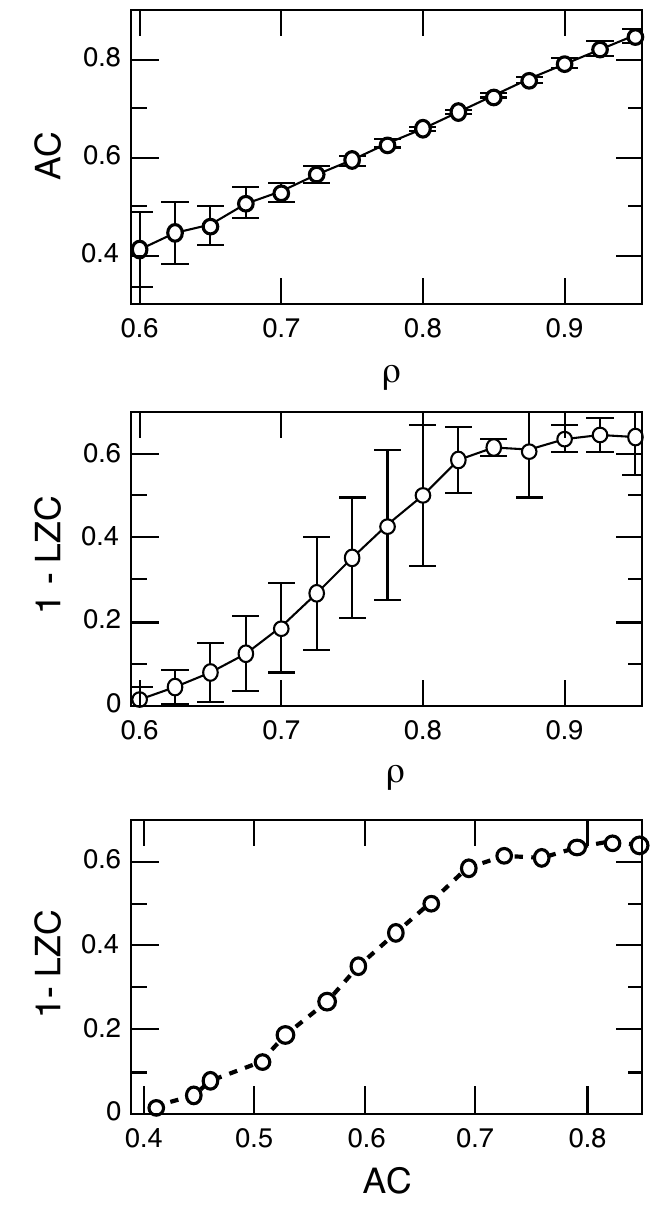}
\caption{1D random Cantor set: Autocorrelation (AC) and Lempel-Ziv complexity (LZC) relation computed numerically. The upper panel shows the Cantor set parameter $P$  and the numerically computed  lag-1 autocorrelation (AC1). The middle panel compares the computed AC1 and the computed Lempel-Ziv complexity (plotted as 1 - LZC) across the range of $\rho$ values. The lower panel depicts the weakly non-linear association between AC and LZC. As $P$ increases, the correlation of the Cantor set increases, approaching a relatively more uniform (i.e., repeating) sequence, which is reflected as expected in the Lempel-Ziv complexity. Parameter $b=1/2$. The data points are means and SDs  from n=64 realizations.}
\label{fig2}
\end{figure}

{\bf \emph{Random Cantor Set:}}
To show that the observed effects are not confined to autoregressive models, we generated additional simulations using random 1-D Cantor sets which permit systematic manipulation of long-range temporal correlations.  A binary sequence was generated from the 1-D process constructed through the following iteration steps \cite{cantor}:
 
\begin{enumerate}
    \item {Initialization.} Start with the unit interval $[0,1]$, 
    denoted $C_0$.
    
    \item  {Subdivision.} At each level $n$, divide every surviving 
    interval of length $L_n$ into $b$ equal sub-intervals of length 
    $L_n / b$.
    
    \item  {Random retention.} Retain each sub-interval independently 
    with probability $P \in (0,1)$. The union of retained intervals forms 
    $C_n$.
    
    \item  {Iteration.} Repeat indefinitely. The limiting set
    \begin{equation}
        C_\infty = \bigcap_{n=0}^{\infty} C_n
    \end{equation}
    is the random Cantor set.
\end{enumerate}

\emph{Fractal Dimension:} For branching number $b$ and retention probability $P$, the almost-sure 
box-counting dimension is
\begin{equation}
    D = \frac{\log(b\,P)}{\log b},
\end{equation}
valid for $P > 1/b$. For $P \le 1/b$ the set is empty (i.e, repeating 0’s) almost surely.

For the binary case $b = 2$:
\begin{equation}
    D = \frac{\log(2P)}{\log 2} = 1 + \frac{\log P}{\log 2}.
\end{equation}

The Hurst exponent is related to the fractal dimension by
\begin{equation}
    H = \frac{D-1}{2}.
\end{equation}

\emph{Autocorrelation function:} For a second-order self-similar process with Hurst exponent $H$, the 
autocorrelation at lag $k$ is
\begin{equation}
    \gamma(k) = \frac{\sigma^2}{2}
    \Bigl[(k+1)^{2H} - 2k^{2H} + (k-1)^{2H}\Bigr],
    \qquad k \ge 1 .
\end{equation}

At lag $1$:
\begin{equation}
    \gamma(1) = \frac{\sigma^2}{2}\bigl(2^{2H} - 2\bigr).
\end{equation}

\emph{First-Lag Autocorrelation Coefficient:} For a point process extracted from the random Cantor set, the lag-1 autocorrelation AC1 has no simple closed form in terms of $P$ alone.  It depends on  the extraction mechanism (thresholding, level crossings, etc.), the correlation structure of the underlying Cantor measure, and of course on the branching number $b$.

The correlation function decays as
\begin{equation}
    \rho(i,i') \sim e^{-|i-i'|/\xi},
    \label{Eq9}
\end{equation}
where $\xi$ is the correlation length, i.e., the average length that separates two uncorrelated samples.

\begin{figure}[h!]
\centering
\includegraphics[width=	.9\linewidth]{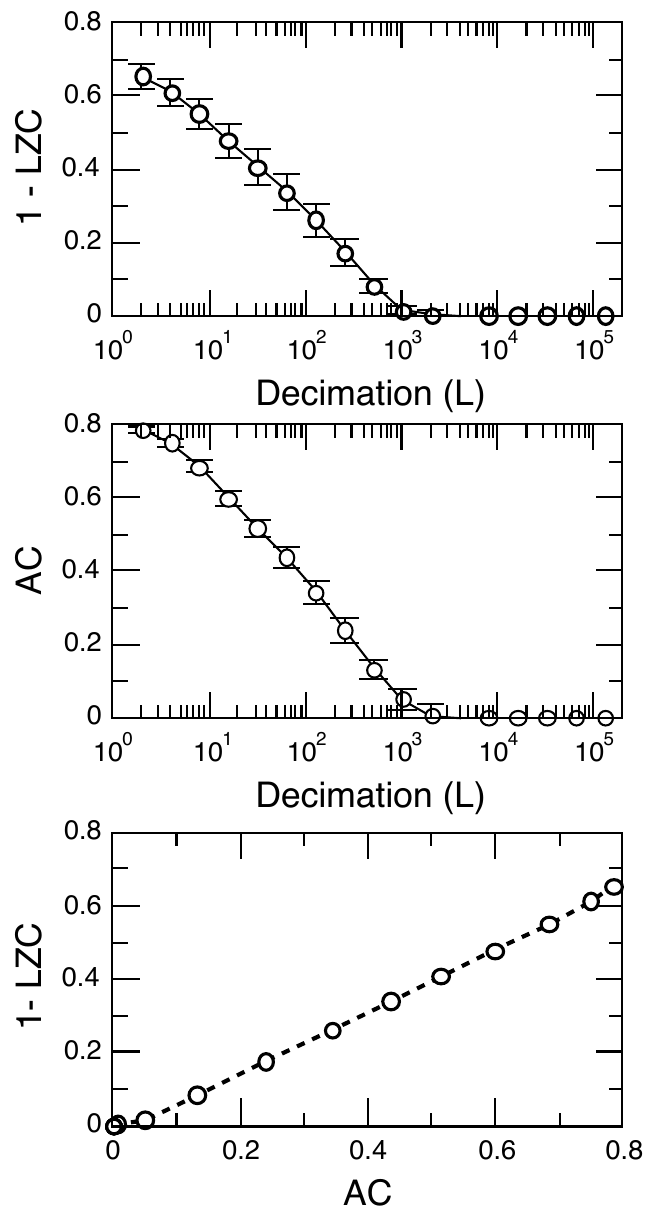}
\caption{Progressive decimation of a correlated 1D random Cantor set: Autocorrelation (AC) and Lempel-Ziv complexity (LZC) relation computed numerically. The top two panels show LZC and AC1 as functions of increasing coarse-graining $L$. The bottom panel shows the same data as the top panels across the coarse-graining range. Notice the close correspondence between the two metrics as decimation proceeds, progressively decorrelating the random Cantor set. The data points are means and SDs from $N=64$ stochastic realizations of the random Cantor set. Parameter values $b=1/2, P=0.9$ . }
\label{fig3}
\end{figure}
We used stochastic realizations of the process to compute metrics similar to those in the Markov chain case. As seen in Figure \ref{fig2}, a monotonic relationship emerges, although LZC exhibits some nonlinear behavior with respect to AC1. Despite that, the results indicate that the association between complexity and temporal dependence extends beyond specific model classes. Thus, although they originate from different theoretical grounds, both measures capture related aspects of temporal organization.

{\bf \emph{Decimation:}} As an additional model, we consider generating a correlated binary sequence and then subjecting it to increasing coarse-graining via decimation. This transformation has been proposed as a simple method for finding the correlation length $\xi$ of equilibrium and non-equilibrium systems, or in fact of any temporal or spatial array (e.g., a sequence or an image). This process yields the correlation length by thinning the data at increasing intervals and recalculating LZC (or AC1). The data becomes incompressible and uncorrelated (i.e., LZC=1 and AC1=0) when the sampling interval exceeds the system’s correlation length. This coarse-graining modifies the sequence’s correlation properties progressively in a predictable way, a procedure we use to further test the consistency of the LZC/AC1 relationship.

The process starts with a correlated binary sequence of length $2^N$ (here we use a 1D random Cantor set with $b=1/2; P=.9$, $N=2^{22}$). In each iteration, we split the sequence into windows of length $2^k$ (k= 2, 4, 8, ...). In each step, the window contents are replaced by their (rounded) mean value. In this way, the sequence progressively converges to a uniform sequence of 0s or 1s (depending on the initial condition). It can be seen that, as expected, the temporal correlation decreases monotonically at each step, eventually reaching zero for a window length that defines the process’s correlation length $\xi$ \cite{decimation,decimation1,decimation2} mentioned in Eq.\ref{Eq9} above.  The same behavior is observed for the LCZ metric, again confirming that both metrics are evaluating similar properties of the time series.
\begin{figure*}[bt!]
\centering
\includegraphics[width=	.8\linewidth]{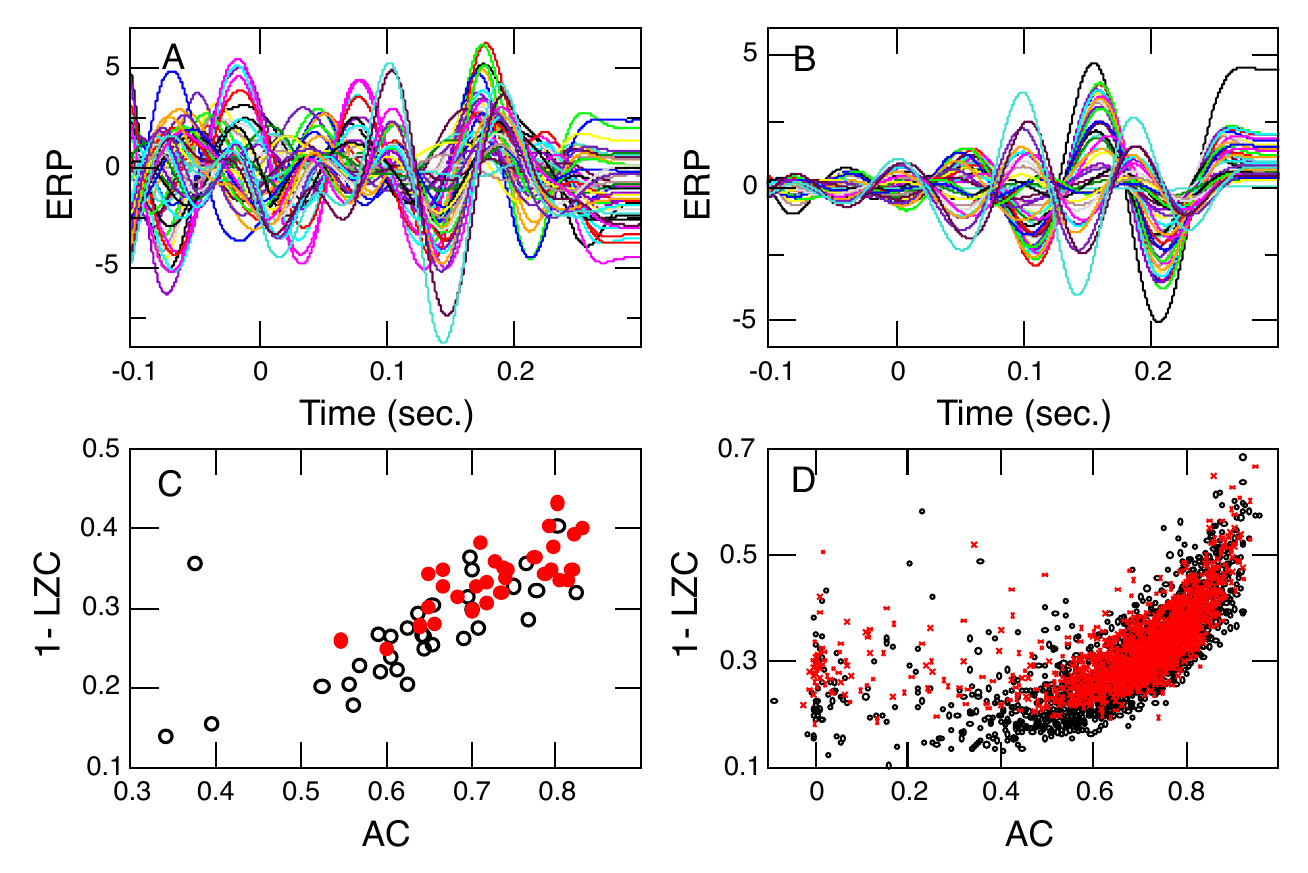}
\caption{Analysis of time series of  EEG $\alpha$-band event-related potentials (ERP) in response to visual imagery (``Happy faces”). Panel A shows typical traces (one for each of 46 electrodes) from a single trial. Time zero corresponds to the time of stimulus presentation.  The results in Panel B correspond to the averages (n= 32 trials) for the same electrodes. 
The computed AC and LZC values shown in Panel C correspond to averages of all electrodes in each single trial, where the empty (black) circles denote the AC/LZC values for the pre-stimulus samples (first 100 msec.) and the filled (red) circles for the last 200 msec. Panel D summarizes the same computations as in Panel C for 51 subjects (n=1632, symbols’ colors as in C). Time series data from \cite{bezhad2026}.}
\label{fig4}
\end{figure*}

\section{Analysis of EEG time series}
To determine whether the relationship observed in synthetic signals is present in physiologically meaningful data, we analyzed two types of human EEG recordings; the first one corresponds to EEG evoked potentials obtained during an emotion-attention task \cite{bezhad2026}, while the second corresponds to a study \cite {EFpaper} of the (lack of) relation between resting-state EEG microstates and executive functions.

We preprocessed the EEG data using standard procedures, including artifact rejection and baseline correction. For the purposes here, we set the filters for the $\alpha$-band (8-13 Hz) and calculate Lempel-Ziv complexity after binarizing the signal using a median threshold. In parallel, the first-lag autocorrelation coefficient was estimated from the timeseries of the same EEG segment. We concatenated individual channels, and presented results across trials and/or participants.

 {\bf \emph{EEG Evoked Potentials: }}The data were acquired as part of a study examining the effects of transcranial direct current stimulation (tDCS) on the processing of emotional face distractors during a color-detection task \cite{bezhad2026}. Participants performed repeated task blocks while EEG activity was continuously recorded and event-related potentials (ERPs) were extracted. This dataset provides a suitable framework for examining temporal predictability because ongoing EEG activity fluctuates substantially across trials despite highly controlled and repeated sensory stimulation. Consequently, individual trials differ in their pre-stimulus dynamical state, yielding a broad range of autocorrelation and complexity values under realistic neurophysiological conditions. 

The EEG dataset analyses revealed a robust relationship between normalized Lempel-Ziv complexity and the first autocorrelation coefficient  (first-lag=0.5 msec.). Epochs with strong temporal persistence consistently produced lower complexity estimates, whereas epochs with randomly fluctuating activity produced higher complexity values.  The data in Fig.\ref{fig4}A show the well-known variability of single-trial EEG traces, revealing a typical response only after averaging many trials, as seen in Panel B.  Despite that, the resulting scatterplots for any given participant (see Fig.\ref{fig4}.C) showed a remarkably stable, approximately linear AC1/LZC relation, both for the pre-stimulus interval (i.e., the initial 100 msec.) as well as for the interval at which the evoked potentials reach their maximum.  The same tendency holds when we pool all participants’ results (see Fig.\ref{fig4}.D). Note that it is not the aim here to interpret the ERPs behavior rather to demonstrate that in a typical ERP experiment the  evoke potential's LZC and AC1 fluctuate in parallel.    

These findings suggest that a large proportion of the variability commonly attributed to changes in complexity may be explained by differences in temporal correlations of the ongoing EEG. In practical terms, whenever the signal exhibits stronger short-term memory, the probability of recurrent symbolic patterns increases, allowing the Lempel-Ziv algorithm to compress the sequence more efficiently and consequently reducing the estimated complexity.

\begin{figure*}[tb!]
\centering
\includegraphics[width=	.8\linewidth]{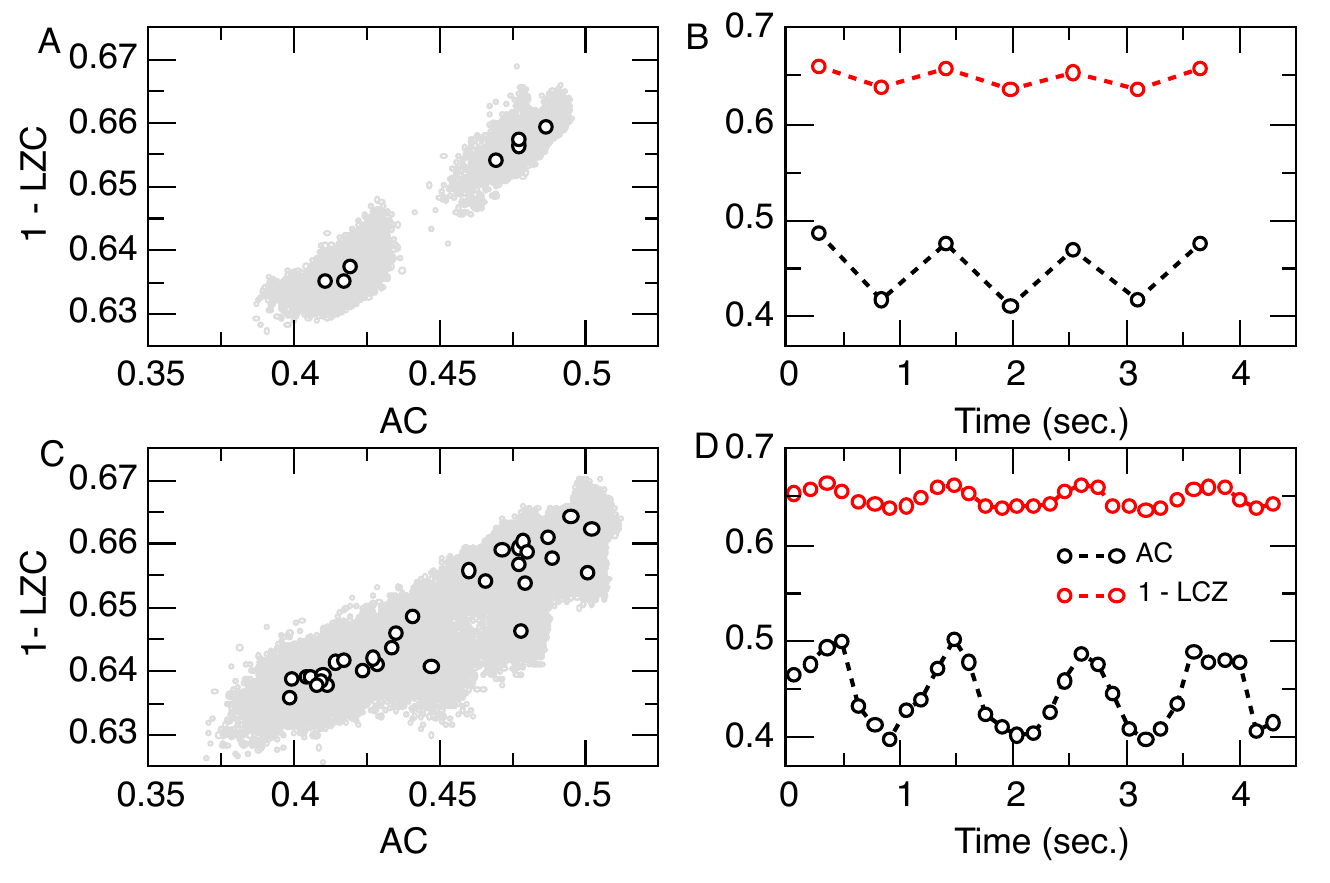}
\caption{Analysis of resting-state EEG recordings (eyes open/closed).  The AC1 (first-lag=2 msec.) and normalized Lempel-Ziv complexity results in Panels A and B correspond to calculations in 5-minute windows (i.e., open or closed conditions), and Panels C and D correspond to calculations on sliding, non-overlapping 75-sec windows. Panels A and C include all channels (pairs of data points), and panels C and D show the average across all channels within the sliding window as a function of time. EEG data from \cite{EFpaper}.}

\label{fig5}
\end{figure*}

{\bf \emph{Resting-state EEG recordings:}}
We now analyze the AC1/LZC relationship in resting-state EEG recordings from a publicly available dataset of subjects instructed to keep their eyes open or closed at 5-minute intervals \cite{EFpaper,EFpaper2} (53 channels, subject ID: ``sub001’’, sampling rate 512 HZ). The eyes open/close maneuver induces cortical changes between more/less synchronous states, which are reflected in the metrics we are interested in. Indeed, the results in Figure \ref{fig5} further validate the AC1/LZC relationship, demonstrating a quasilinear relation between the two metrics. Following the same procedure as in the previous section, we computed the two metrics for each electrode (in this study, $n=53$). For the results in Fig.\ref{fig5}A,B, the calculations were done separately for eyes open or closed, while for the results in Fig.\ref{fig5}C,D, at sliding windows of 75 sec.   As expected, the eyes-open/closed conditions differ, resulting in two well-separated groups of points (Panel A). One cluster corresponds to eyes open, with relatively low AC1 and high LZC, and vice versa for eyes closed. Furthermore, we show the AC1/LZC dependency over time by computing these quantities within a relatively short (75 secs.) sliding window, as shown in Panels C and D. 

The relationship between correlations and complexity matters beyond simple time-series analysis. Previous work has shown that brain function operates near a phase transition across scales \cite{chialvo2010,decofmri}. Of note is that among the properties that complex systems exhibit near a critical point, the increase of AC1 is ubiquitous  \cite{refsAC1_criticality}. As a complex system approaches the edge of synchronization, its average activity becomes more correlated both in space (with other constituents) and in time (so-called critical slowing down). This is not surprising, since to become ordered, the system must increase correlations. What is non-trivial it is the fact that such a feature is independent of the details of the system; the same increase in AC1 can be seen in a flock of birds, in a magnetic spin collective, or in a population of neurons of a spiking neural network, as demonstrated recently in \cite{refsAC1_criticality}. In addition, under these circumstances, increases in AC1 are expected to be accompanied by similar increases in the value of the pair correlations (i.e., cross correlations (CC) or also dubbed ``functional connectivity’’). 

These considerations suggest that AC1 (and LZC) can track spontaneous arousal fluctuations, assuming they reflect fluctuations in the distance to the system critical point. In that direction, some evidence was recently presented in Galbardon et al. \cite{Galbardon}.

\section{Discussion}
The present analyses demonstrate a robust and systematic association between normalized Lempel-Ziv complexity (LZC) and temporal autocorrelation in both synthetic signals and physiological EEG recordings. Across a broad range of signal conditions, increased temporal persistence was consistently associated with lower LZC values, whereas signals characterized by reduced predictability exhibited higher complexity estimates. Although the relationship varied across signal classes, the overall pattern remained remarkably stable, suggesting that temporal dependence is a major determinant of LZC in EEG data.

From a theoretical perspective, this finding is not surprising. The Lempel-Ziv algorithm quantifies the rate at which new patterns appear within a symbolic sequence. Signals with recurrent temporal structures generate more repeated patterns and are therefore more compressible, yielding lower complexity values. Temporal autocorrelation reflects precisely this tendency for successive samples to resemble one another. Consequently, both measures reflect a common underlying property of the signal: temporal predictability. In this sense, temporal autocorrelation may be viewed as one of the principal factors governing the compressibility from which LZC is derived.
Importantly, our results do not imply that LZC and autocorrelation are mathematically equivalent.   Higher-order correlations are not represented by a single autocorrelation coefficient, and signals  may differ in other aspects of temporal organization. Nevertheless, the observed relationship indicates that a substantial proportion of the variability commonly attributed to the more often used complexity measures can be explained by relatively simple temporal dependencies.

This observation has implications for the physiological interpretation of EEG complexity. Changes in autocorrelation often relate directly to known neurophysiological mechanisms. Increased temporal persistence may arise from stronger rhythmic activity, greater oscillatory stability, enhanced synchronization among neuronal populations, or slower recovery of cortical networks following spontaneous fluctuations. Such mechanisms are readily interpretable within established frameworks of systems neuroscience and electrophysiology.
In contrast, changes in LZC are frequently interpreted in terms of broad concepts such as neural complexity, information richness, dynamical repertoire, or conscious content. While these interpretations may be useful at a phenomenological level, they are inherently indirect because the algorithm itself does not quantify any of these physiological processes. Rather, it quantifies the compressibility of a symbolic sequence. Consequently, similar changes in LZC may arise from very different underlying mechanisms, including alterations in oscillatory activity, spectral composition, synchronization, signal-to-noise ratio, or temporal dependence. As a result, the physiological meaning of a given complexity change is often ambiguous.

The practical implications are equally relevant. Estimating autocorrelation is computationally trivial, requires no symbolization or thresholding procedures, and offers a direct, intuitive interpretation as a measure of temporal memory. By contrast, computing LZC requires signal discretization, depends on parameter choices such as sequence length and symbolization strategy, and yields an index whose physiological interpretation is far less transparent. If both measures are found to track the same underlying property of the EEG, the additional complexity of the LZC framework may not always be justified.
This does not diminish the potential usefulness of LZC. Complexity metrics remain valuable as compact descriptors of signal organization, and under certain conditions they may capture aspects of temporal structure not reflected in conventional linear statistics. However, before attributing physiological significance to changes in Lempel-Ziv complexity, it is important to establish whether simpler signal properties can explain these changes. In this regard, temporal autocorrelation should be considered a natural reference measure against which a potential added value of complexity metrics can be evaluated. 

In terms of interpretation and mechanisms, we noted the established observation that temporal correlation universally increases as a complex system loses stability near a transition point \cite{chialvo2010,refsAC1_criticality,grigera,decofmri}. Because the critical state also predicts a set of testable dynamical properties, this feature may be useful when the objective is to mechanistically interpret spontaneous or induced (drug, arousal state, etc.) fluctuations of AC1 (and LZC) in a given neurophysiological setting.  

Taken together, the present results suggest that, as expected from its construction, Lempel-Ziv complexity reflects the signal’s temporal correlations. Consequently, its use for typical EEG recordings may not capture more information than standard correlation approaches. Autocorrelation provides this information through a simpler computational procedure and with a more direct physiological interpretation. These results therefore suggest that studies using complexity metrics should routinely assess their relationship with autocorrelation and other basic descriptors of temporal structure before assigning additional functional significance to observed changes in EEG complexity.

\section {Acknowledgments} {PM is supported by AEI and ERDF (MICIU/AEI/10.13039/501100011033, grant PID2022-140561NB-I00). IC was supported by AEI and ERDF (grant PID2021-125534OB-I00). DRC thanks the UIB’s Visiting Professorship Program under which these studies were conducted. JRS was partially supported by the Ministerio de Ciencia e Innovación and by European Union NextGenerationEU/PTR. under grant contracts PID2020-120075RB-I00 and PID2024-158410OB-I00 funded by MCIN/ AEI/10.13039/501100011033.}

\end{document}